\documentclass[11pt]{article}

\usepackage[T1]{fontenc}
\usepackage[utf8]{inputenc}
\usepackage[margin=1in]{geometry}
\usepackage{booktabs}
\usepackage{graphicx}
\usepackage[hidelinks]{hyperref}
\usepackage{xurl}
\usepackage{array}
\usepackage{longtable}
\usepackage{caption}
\usepackage{amsmath}
\usepackage{amssymb}
\usepackage{float}

\begin{document}

\title{Source-grounded biomedical evidence records from full-text disease-focused literature}

\author{%
Chang Zong$^{1,*}$, Jinyu Chen$^{1}$, Sicheng Lv$^{1}$, Si-tu Xue$^{2}$, Huilin Zheng$^{3}$, Jian Wan$^{3}$, Lei Zhang$^{1,*}$\\[0.6em]
\small \parbox{0.92\textwidth}{\centering
$^{1}$School of Computer Science and Technology, Zhejiang University of Science and Technology,\\
Hangzhou, China\\
$^{2}$Institute of Medicinal Biotechnology, Chinese Academy of Medical Sciences \&\\
Peking Union Medical College, Beijing, China\\
$^{3}$Zhejiang Key Laboratory of Biomedical Intelligent Computing Technology, Hangzhou, China}\\[0.4em]
\small *Correspondence: \href{mailto:zongchang@zust.edu.cn}{zongchang@zust.edu.cn}, \href{mailto:leizhang@zust.edu.cn}{leizhang@zust.edu.cn}
}

\date{}
\maketitle

\begin{abstract}
Potentially reusable biomedical findings are often embedded in full-text articles and incompletely represented in article-level metadata or compact knowledge-graph triples. Here we present EvidenceNet, a public collection of source-grounded biomedical evidence records derived from full-text literature on diseases including hepatocellular carcinoma, colorectal cancer and systemic lupus erythematosus. The release contains 18,755 record-level JSON evidence records and derived companion graph files with 25,465 nodes and 110,762 edges. Each record preserves article provenance, a supporting source-text passage, structured study context, biomedical entities, study design, clinical stage, quantitative attributes and links to normalized entities or related evidence records. The dataset is generated through a standardized full-text processing and record-construction workflow and is technically validated for provenance, record-graph consistency, field completeness, metadata quality and grounded entity linkage. EvidenceNet supports literature screening, evidence retrieval, evidence synthesis and related biomedical informatics applications across disease-focused corpora.
\end{abstract}

\section{Background \& Summary}

Evidence-based medicine depends on connecting biomedical decisions to the best available evidence rather than to isolated claims\cite{sackett_ebm}. In practice, however, much of the evidence needed for reuse is dispersed across full-text articles. A reported observation may depend on the study population or model system, intervention, comparator, outcome, experimental context, source passage and quantitative support. These details are often difficult to recover from article-level metadata or from compact entity-relation databases alone. This limitation matters more as biomedical literature is increasingly reused by retrieval systems, literature-screening tools, knowledge graphs and language-model applications, all of which require data that can be traced back to specific source text rather than only to a citation.

Article-level bibliographic databases are indispensable for discovery, but they do not usually indicate which passage supports a reusable biomedical observation. Compact structured resources such as knowledge-graph triples support large-scale navigation across concepts and relations, yet they often omit the local source passage, study context and quantitative detail needed for close evidence inspection. Expert syntheses such as systematic reviews and evidence maps provide carefully assessed summaries, but they are expensive to prepare and are usually scoped to narrow clinical questions. Between article-level discovery data and more condensed structured or expert-synthesized representations lies a practical intermediate form: machine-readable records that preserve the local evidence context of reported observations. Such records are useful for literature screening, evidence synthesis and structured evidence review. The population, intervention, comparison and outcome (PICO) framework\cite{richardson_pico,jbi_manual} provides a familiar way to organize clinical and translational questions, while GRADE-like appraisal\cite{guyatt_grade} highlights the importance of study design, statistical support and certainty when evidence is reused. EvidenceNet adapts these ideas as record fields and quality-related annotations while preserving the supporting source passage.

Structured biomedical resources make large-scale biomedical knowledge more accessible. Hetionet\cite{hetionet} integrates biomedical entities for drug repurposing, while PrimeKG\cite{primekg} combines multiple resources into a broad precision-medicine graph. Disease- and domain-specific knowledge graphs such as HALD\cite{hald} and PubMed-derived knowledge graphs\cite{pubmed_kg,pubmed_kg2} show that graph-structured biomedical data can support navigation across heterogeneous sources. TarKG\cite{tarkg} and Open Targets\cite{open_targets} illustrate the value of target-oriented data integration and harmonization, and LitCovid\cite{litcovid} shows how a disease-focused literature resource can support literature surveillance and text mining during rapid publication growth. These resources are valuable, but they usually emphasize entities, concept annotations, bibliographic links or typed relations. A user who wants to examine the evidence behind a disease-specific claim still often needs to return to the full text to determine the population, model system, intervention, comparator, outcome and strength of support.

Biomedical text-mining resources provide another relevant comparison. SemMedDB\cite{semmeddb}, PubTator Central\cite{pubtator}, BioRED\cite{biored} and text-derived biomedical relationship networks\cite{percha_network} show the value of extracting concepts, relations and semantic predications from abstracts or full text at scale, while recent surveys highlight the continued importance of integrating domain knowledge into biomedical text analysis workflows\cite{cai_biomedical_text_survey}. Biomedical word embeddings and language models, including BioWordVec\cite{biowordvec}, BioBERT\cite{biobert}, PubMedBERT\cite{pubmedbert} and BioGPT\cite{biogpt}, further show that curated biomedical text resources remain central to information extraction and literature analysis\cite{pubmedqa}. Recent \textit{Scientific Data} resources extend this landscape with reusable biomedical corpora and evaluation sets, including BioASQ-QA\cite{bioasq_qa_sdata}, a dataset for plain-language adaptation of biomedical abstracts\cite{plaba_sdata} and a benchmark dataset for contextualized biomedical concept representations\cite{biomed_concept_repr_sdata}. Recent foundation-model studies in medicine renew interest in reliable biomedical corpora for evidence retrieval and verification\cite{moor_foundation,singhal_medpalm,multimodal_biomed_ai}. In this setting, a source-grounded evidence record offers a practical unit between unstructured full text and highly compressed knowledge-graph triples. Table~\ref{tab:resource_positioning} summarizes these differences in data organization and reuse.

\begin{table}[ht]
\centering
\caption{Positioning of EvidenceNet relative to article-level literature resources and compact biomedical knowledge graphs.}
\label{tab:resource_positioning}
\small
\renewcommand{\arraystretch}{1.18}
\begin{tabular}{>{\raggedright\arraybackslash}m{0.2\textwidth}|>{\raggedright\arraybackslash}m{0.24\textwidth} >{\raggedright\arraybackslash}m{0.24\textwidth} >{\raggedright\arraybackslash}m{0.24\textwidth}}
\toprule
Aspect & \shortstack[l]{Article-level literature} & \shortstack[l]{Compact knowledge graph} & \shortstack[l]{EvidenceNet release} \\
\midrule
\shortstack[l]{Primary unit} & \shortstack[l]{Article abstract} & \shortstack[l]{Triple typed edge} & \shortstack[l]{Evidence record} \\
Provenance & \shortstack[l]{Article-level metadata} & \shortstack[l]{Citation-level} & \shortstack[l]{Article metadata\\+ source span} \\
\shortstack[l]{Study context} & \shortstack[l]{Free-text narrative} & \shortstack[l]{Sparse or compressed} & \shortstack[l]{Structured context fields} \\
\shortstack[l]{Quantitative detail} & \shortstack[l]{Reported in text} & \shortstack[l]{Rarely retained} & \shortstack[l]{Captured when reported} \\
\shortstack[l]{Typical use} & \shortstack[l]{Search\\Reading\\Screening} & \shortstack[l]{Navigation\\Integration\\Graph mining} & \shortstack[l]{Retrieval\\Synthesis\\Structural mining} \\
\bottomrule
\end{tabular}
\end{table}

Within the categories in Table~\ref{tab:resource_positioning}, integrated resources such as PrimeKG and PubMed-derived knowledge graphs\cite{primekg,pubmed_kg,pubmed_kg2} emphasize connected entities, typed links and aggregate coverage, whereas text-mining resources such as SemMedDB and BioRED\cite{semmeddb,biored} emphasize semantic predications or annotated mention-level relations. EvidenceNet is also literature-derived, but its deposited data object is a source-grounded evidence record that keeps the supporting passage, study-context fields and quantitative attributes together as a reusable unit.

The current release covers hepatocellular carcinoma (HCC), colorectal cancer (CRC) and systemic lupus erythematosus (SLE). HCC and CRC contribute two cancer domains with large and clinically diverse literatures, including first-line immunotherapy combinations for unresectable HCC\cite{global_cancer,finn_imbrave150,qin_cares310}, circulating-tumour-DNA-guided adjuvant decisions in stage II colon cancer\cite{tie_dynamic} and PD-1 blockade in mismatch-repair-deficient locally advanced rectal cancer\cite{cercek_rectal_pd1}. SLE extends the same schema to an autoimmune disease domain with distinct clinical and molecular evidence patterns, including recent B-cell-targeted therapy studies in active lupus nephritis\cite{furie_obinutuzumab_ln}. Together, these three subsets provide aligned record files, graph views, validation samples, metadata-correction records and checksums under a common quality-control workflow.

Each EvidenceNet record is linked to a supporting source-text passage and preserves article metadata, structured study context, extracted entities, study design, clinical stage, quantitative attributes and quality-related fields. Companion graph files are generated from the same records, while the record files remain the most detailed representation of the deposited data. In total, the release contains 18,755 record-level JSON objects and companion graph files with 25,465 nodes and 110,762 edges. The deposited package also includes schemas, prompt templates, validation files and quality-control artefacts, following the FAIR principle that reusable data should be findable, accessible, interoperable and reusable\cite{fair}. This manuscript describes the released files, the steps used to create them and the technical validation applied to provenance, schema consistency, field completeness, entity grounding and source-record lookup.

\section{Methods}

Figure~\ref{fig:workflow} summarizes the dataset-construction workflow. Starting from 1,334 disease-specific source PDFs represented in the released record provenance fields, the workflow proceeds through corpus assembly, full-text parsing, evidence-record extraction, entity normalization, duplicate handling, evidence-to-evidence linking and release assembly. The deposited package contains record-level JSON files as the primary data objects, together with companion graph files and supporting documentation for validation and reuse.

\subsection{Input literature and eligibility criteria}

For each disease domain, we assemble a corpus of full-text biomedical PDF articles from disease-focused literature searches and locally curated full-text collections. Candidate papers are first identified from PubMed (\url{https://pubmed.ncbi.nlm.nih.gov/}) using the target disease names and common synonyms. The search terms used for corpus assembly include \emph{hepatocellular carcinoma}, \emph{HCC} and \emph{liver cancer} for HCC; \emph{colorectal cancer}, \emph{colon cancer}, \emph{rectal cancer} and \emph{CRC} for CRC; and \emph{systemic lupus erythematosus}, \emph{SLE}, \emph{lupus nephritis} and \emph{lupus} for SLE. Accessible full-text PDFs are then downloaded for local processing and organized by disease into fixed release subsets. Corpus assembly, article inclusion and metadata QC for the deposited release are frozen in May 2026 so that the release boundary remains stable. The released records are grounded in 470 unique HCC source PDFs, 472 CRC source PDFs and 392 SLE source PDFs, as identified from the deposited \texttt{source.pdf\_path} provenance fields (Table~\ref{tab:input_corpus}). A PDF contributes records when its full text contains disease-relevant experimental, clinical or translational evidence, or systematic reviews with extractable disease-specific findings, and at least one observation can be grounded in a supporting source-text span. Files that yield only bibliographic information, cannot be parsed into usable text, or contain only general background discussion are not used for record creation.

\subsection{Disease-domain scope and corpus boundary}

The corpus boundaries for this release are defined by three target diseases: hepatocellular carcinoma (HCC), colorectal cancer (CRC) and systemic lupus erythematosus (SLE). HCC and CRC are selected as two clinically important solid-tumour domains with substantial therapeutic, biomarker and mechanistic literature, allowing the release to cover heterogeneous evidence types ranging from molecular studies and preclinical experiments to clinical cohorts and treatment-response studies. SLE is added to extend the same record schema to a systemic autoimmune disease setting with different evidence patterns, including immune dysregulation, nephritis, autoantibodies and longitudinal treatment-response analyses. Together, these three domains provide a cross-domain test of data quality and schema coverage while remaining narrow enough for aligned corpus assembly, metadata checking, validation sampling and release documentation. After metadata QC, the released subsets cover source articles published between 2000 and 2026. Additional diseases can be incorporated in later dataset versions when the same level of record-level validation and release documentation is completed.

For each disease, the processing log records the PDF files that contribute to the released records. Bibliographic metadata are associated with each article where available, including title, DOI, journal, publication year, authors, citation count, impact factor and journal quartile. DOI-based metadata recovery is attempted first and title-based recovery is used when DOI information is unavailable or incomplete. The deposited \texttt{source} metadata within the record files preserve the exact retained article set, including article titles and DOIs where available, so the input literature used for record construction can be identified from the released files themselves. Table~\ref{tab:input_corpus} therefore emphasizes record-linked source PDFs, retained evidence records and unique article counts, which are more directly comparable across the three released subsets than the size of the local candidate PDF collections.

\begin{figure}[!tbp]
\centering
\includegraphics[width=0.95\textwidth]{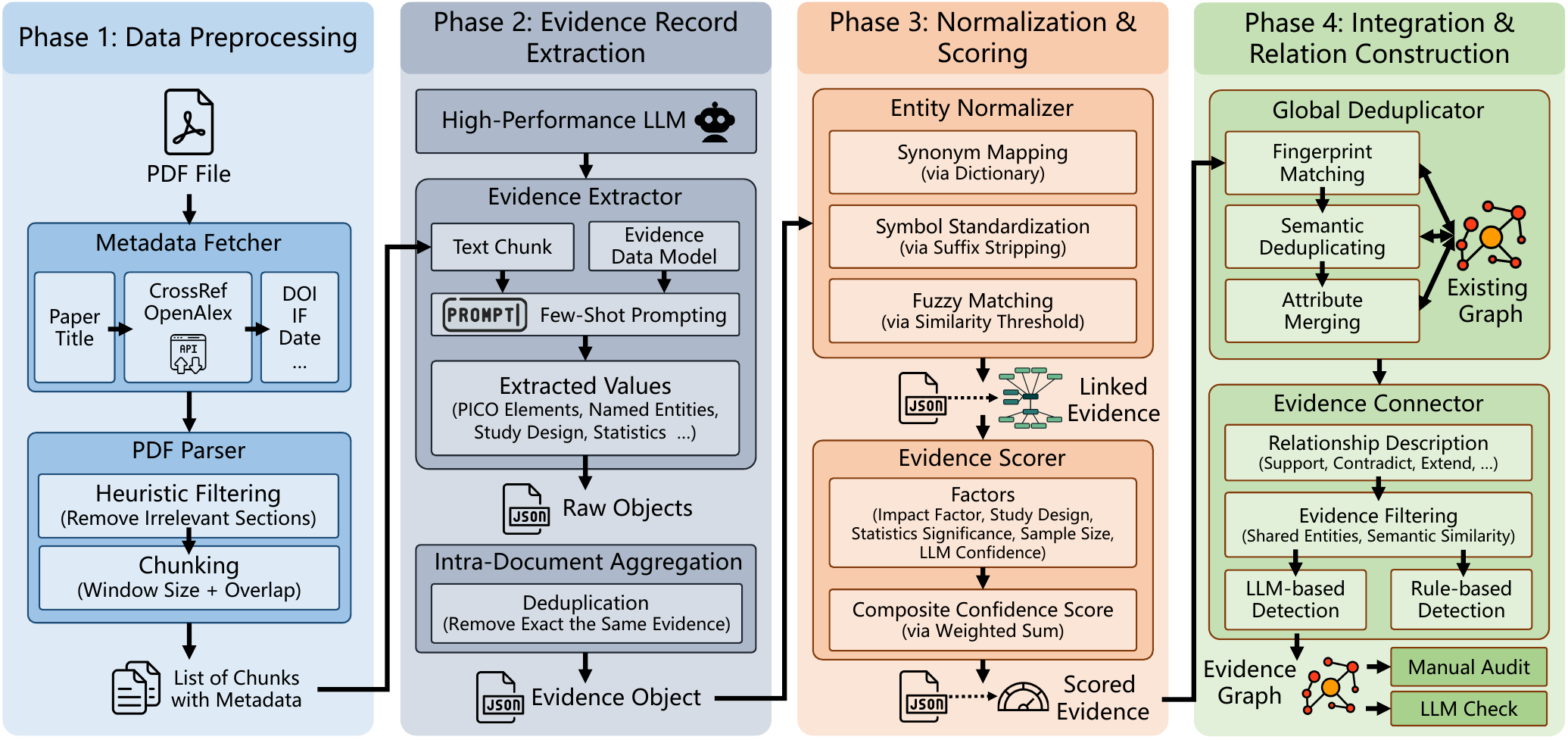}
\caption{Dataset construction workflow. Disease-specific full-text PDF collections are parsed into text, filtered into extractable sections, converted into source-grounded evidence records, normalized to biomedical entities where possible, checked for duplicate or related records and exported as record-level JSON files with derived graph files.}
\label{fig:workflow}
\end{figure}

\begin{table}[!tbp]
\centering
\caption{Input-literature and record-construction summary.}
\label{tab:input_corpus}
\small
\begin{tabular}{lrrr}
\toprule
Measure & HCC & CRC & SLE \\
\midrule
Unique source PDFs linked in records & 470 & 472 & 392 \\
Evidence records retained & 7872 & 6622 & 4261 \\
Records per linked source PDF & 16.8 & 14.0 & 10.9 \\
Unique article titles & 457 & 452 & 382 \\
\bottomrule
\end{tabular}
\end{table}

\subsection{Full-text parsing and section filtering}

Each PDF is parsed into plain text and, when possible, segmented into canonical scientific sections. Reference sections, acknowledgements, conflict-of-interest statements and other low-information regions are excluded before extraction because most evidence-bearing content is concentrated elsewhere in the article. Each retained section is then split into overlapping segments of up to 3,000 characters with 300-character overlap, so adjacent segments begin 2,700 characters apart. Chunks shorter than 500 characters are not submitted for extraction because they often contain headings, author information or fragments with insufficient context. This segmentation retains enough local context for record extraction while reducing boundary effects when an observation crosses adjacent text segments.

\subsection{Evidence record extraction}

Retained chunks are processed with a deterministic extraction protocol with temperature set to zero. In the released workflow, chunks unlikely to contain experimental findings are filtered before full extraction, the primary extraction model is OpenAI \nolinkurl{GPT-5.1} and lighter auxiliary model calls use OpenAI \nolinkurl{GPT-5-mini} for aggregation or later verification tasks, as recorded in the archived configuration bundled with the code release. Biomedical text-mining systems have long used named-entity recognition, relation extraction and language representations to structure literature at scale\cite{pubtator,semmeddb,biobert}. Here, the extraction schema is centered on source-grounded evidence records rather than isolated entity mentions. It requests experimentally grounded records, excludes background statements, review-style summaries and unsupported inference, and requires each candidate record to preserve a supporting source-text span. Where available, the extracted fields include study object, intervention, comparison and outcome fields consistent with the PICO framing used in evidence-based clinical questions\cite{richardson_pico,jbi_manual}, together with biological mechanism, phenotype, study design, clinical stage and quantitative attributes such as sample size, \(p\)-value, effect size, confidence interval or fold change. Missing quantitative fields are left empty rather than inferred.

Chunk-level outputs from the same article are then aggregated at the document level. Exact duplicates are removed when the same finding is repeated across adjacent sections, whereas distinct experiments are retained as separate records when they differ in study object, intervention, comparison, outcome or biological context. Across the three released subsets, this aggregation yields 18,755 retained evidence records. This conservative aggregation keeps the deposited data at the level of reusable observations rather than article-level summaries and matches the record-construction workflow shown in Figure~\ref{fig:workflow}. Because hosted LLM endpoints may evolve over time, the deposited JSON files and archived prompts should be treated as the authoritative release products for reuse, while a future rerun should be regarded as a reconstruction rather than an identical reissue of the same dataset.

Each record retains the record identifier, source metadata, structured study context, extracted entities, study-design and clinical-stage annotations, quantitative attributes, source-text span, and quality-related or relation annotations. The identifier is retained across the record and graph files so that linked graph nodes can be traced back to the full source-grounded record.

\subsection{Entity normalization and evidence-quality scoring}

Biomedical entities extracted in each evidence record are normalized to TarKG-compatible entity identifiers where possible\cite{tarkg}. The matching procedure uses exact matching, curated alias expansion, gene or protein symbol normalization and precision-oriented fuzzy matching, with unresolved mentions left in their original article form. Entity classes in the released records include Gene, Drug or Compound, Disease, Phenotype and Pathway. These classes align with common biomedical resource layers represented in public data infrastructures such as EMBL-EBI resources\cite{ebi_2023}, UniProt\cite{uniprot_2023} and STRING\cite{string_2021}, but the deposited records retain the original article mention alongside any normalized identifier. This preserves abbreviations, historical gene symbols, drug synonyms and disease-specific phenotype names that remain useful during manual review even when a normalized identifier is unavailable.

For candidate fuzzy matches, string similarity is computed after lower-casing, punctuation normalization and removal of gene-organism suffixes where appropriate. Exact and alias matches are preferred over fuzzy matches, and unresolved mentions are retained without a normalized identifier.

Each record also contains a quality-score object that summarizes study design, source impact, reported statistical support, sample size and extraction confidence for filtering and prioritization during reuse. These fields are inspired by evidence-appraisal practice but are not formal GRADE ratings\cite{guyatt_grade}. The released files retain both component values and a composite score, together with grade levels A, B, C and D derived from fixed score intervals. Within the release, these score and grade fields function as filtering annotations and are intended to be read alongside the deposited source text and metadata.

\subsection{Duplicate handling and relation construction}

Within-document and cross-document duplicate handling uses evidence fingerprints and semantic similarity. Candidate duplicates are reviewed for fusion when their identifying fields indicate the same underlying observation or when their cosine similarity, computed with \nolinkurl{sentence-transformers/all-MiniLM-L6-v2}, exceeds the release threshold of 0.92. When highly overlapping records are merged, the higher-quality record is retained as the canonical version and the retained record stores version information together with identifiers of absorbed records so that provenance remains inspectable.

Evidence-to-evidence relations then connect records with annotation labels such as SUPPORTS, EXTENDS, REFINES, CONTRADICTS, CAUSAL\_CHAIN and REPLICATES. Candidate record pairs are proposed from semantic similarity together with overlap in biomedical entities and study context, using a release threshold of 0.75 for relation candidacy and LLM-based second-pass verification at the configured high-similarity checkpoints. Accepted relation labels are stored as annotations in the released files. Evidence-to-entity links connect records with normalized biomedical entities, allowing the companion graph files to preserve both evidence--entity and evidence--evidence connectivity while remaining traceable to the record layer.

\subsection{Release assembly and versioning}

After record construction, scoring, duplicate handling and relation linking, the resulting disease-specific record files are organized as versioned JSON release units for deposition. Companion graph files are generated from the same record identifiers so that evidence nodes in the graph layer can be traced back to the corresponding record-level files. For the deposited release, these primary record files and companion graph files are accompanied by JSON schemas, a field dictionary, validation outputs, metadata-QC records, prompt templates and SHA-256 checksums that document file structure and quality control. The final package contains 18,755 evidence records together with 25,465 graph nodes and 110,762 graph edges across the three disease subsets.

Before deposition, each disease subset is checked so that the record files, graph files and accompanying release documents remain internally aligned. The manuscript therefore describes a three-domain release with consistent documentation across all subsets.

\section{Data Record}

\begin{figure}[!tbp]
\centering
\includegraphics[width=\textwidth,height=0.78\textheight,keepaspectratio]{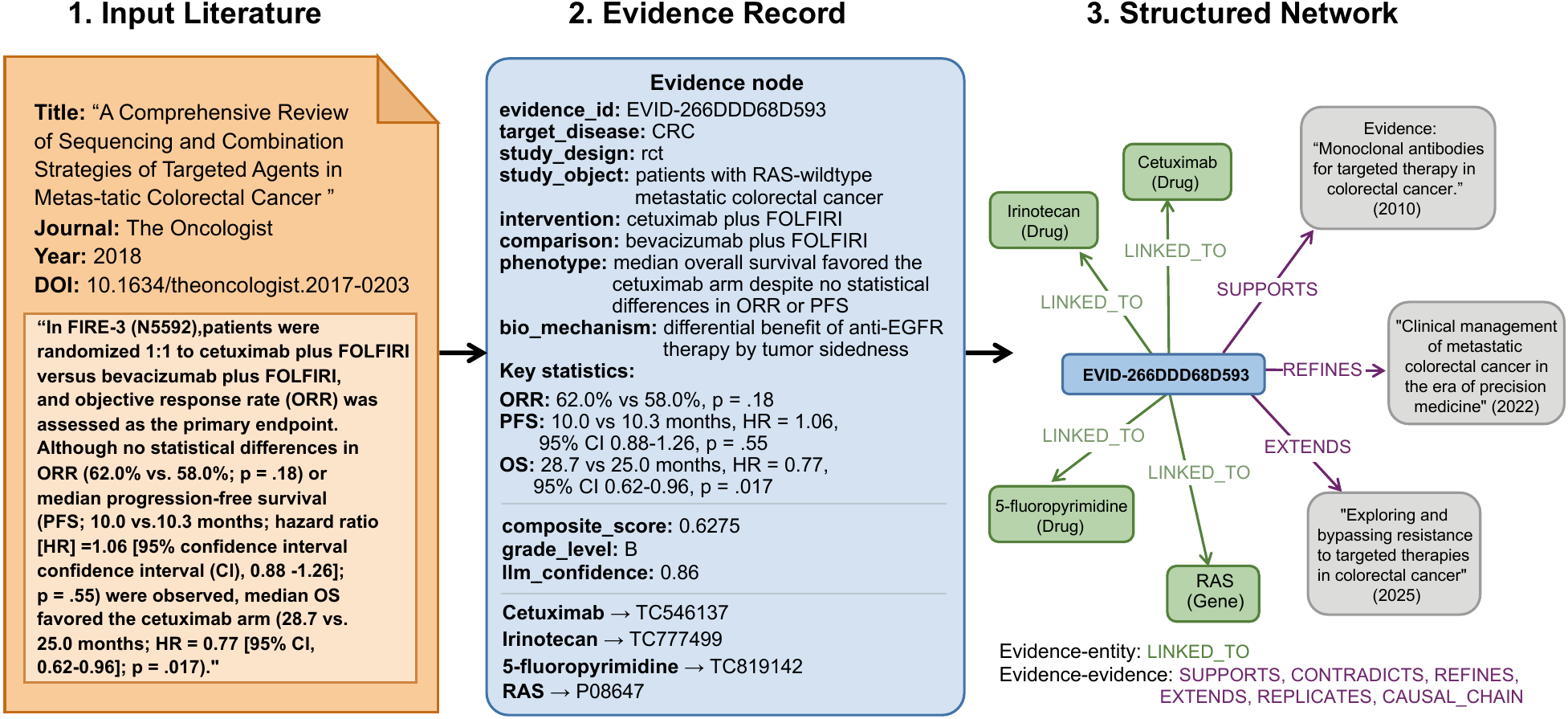}
\caption{Representative transformation from full-text literature to a structured evidence record and its companion relation view. The left panel shows preserved article-level provenance and supporting source text. The middle panel shows the corresponding structured evidence object with provenance, context, finding, entity and quality fields. The right panel shows how the same \texttt{evidence\_id} is linked to normalized entities and related records in the companion graph files.}
\label{fig:method_schema_overview}
\end{figure}

\begin{figure}[!tbp]
\centering
\includegraphics[width=\textwidth,height=0.75\textheight,keepaspectratio]{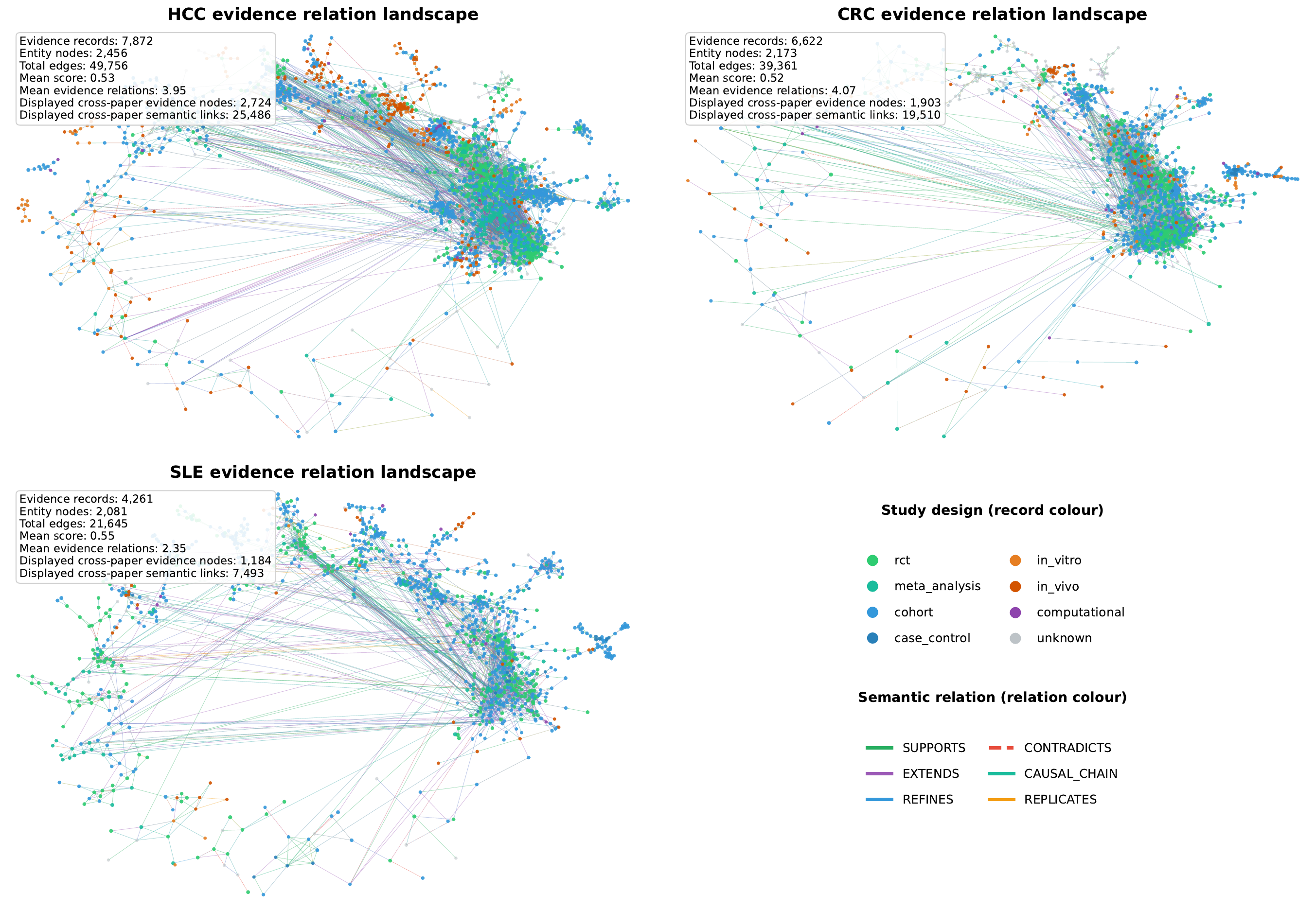}
\caption{Score-filtered cross-paper semantic display views of the released disease-specific companion graph files. Each panel shows a disease-specific display subgraph derived from the corresponding released graph JSON file after restricting to cross-paper evidence-to-evidence relations above the visualization threshold. Record colour indicates study design and relation colour indicates semantic relation type.}
\label{fig:data_overview_global}
\end{figure}

The dataset is deposited at Figshare as three disease-specific release units for HCC, CRC and SLE. The record-level JSON files are the primary deposited data objects, while the graph JSON files provide companion views built from the same evidence identifiers. This organization aligns the release with disease-specific corpus boundaries, preserves common field definitions across the three subsets and allows users to combine subsets for broader cross-disease analyses. Table~\ref{tab:file_manifest} first lists the exact released file groups. Figure~\ref{fig:method_schema_overview} and Table~\ref{tab:record_fields} next describe the structure of a representative record and its main fields. Figure~\ref{fig:data_overview_global} then illustrates the structural form of the companion graph files.

\begin{longtable}{>{\raggedright\arraybackslash}m{5.9cm}>{\raggedright\arraybackslash}m{2.2cm}>{\raggedright\arraybackslash}m{7.4cm}}
\caption{Main deposited file groups and their roles.}
\label{tab:file_manifest}\\
\toprule
File group & Format & Contents\\
\midrule
\endfirsthead
\toprule
File group & Format & Contents\\
\midrule
\endhead
\bottomrule
\endfoot
\shortstack[l]{\texttt{records/}\\\texttt{evidence\_records\_\{disease\}.json}} & JSON & Primary evidence-record files for HCC, CRC and SLE, each keyed by \texttt{evidence\_id}.\\
\shortstack[l]{\texttt{graphs/}\\\texttt{evidence\_graph\_\{disease\}.json}} & JSON & Derived NetworkX node-link graph files for the corresponding disease subsets.\\
\texttt{schema/} & JSON, TSV & Record schema, graph schema and field dictionary for released file interpretation.\\
\texttt{validation/} & JSON, CSV & Audit samples, integrity checks, external-concordance summaries, record-lookup outputs and metadata-QC records.\\
\texttt{prompts/} & TXT & Prompt templates used during extraction, aggregation and relation verification.\\
\texttt{checksums\_sha256.txt} & TXT & SHA-256 checksums for release-level integrity checking.\\
\end{longtable}

Each record-level JSON file is keyed by \texttt{evidence\_id}. The corresponding value is a nested evidence object whose main fields are listed in Table~\ref{tab:record_fields}. Figure~\ref{fig:method_schema_overview} shows the same record logic through a representative example, following one source-grounded finding from full-text provenance to structured evidence fields and then to its companion relation view. This record-level representation is the primary form of the dataset when a user needs to inspect the source passage, article metadata or field-level annotations behind a biomedical observation.

\begin{longtable}{>{\raggedright\arraybackslash}m{4cm}>{\raggedright\arraybackslash}m{11cm}}
\caption{Main fields in the primary evidence-record JSON files.}
\label{tab:record_fields}\\
\toprule
Field & Description\\
\midrule
\endfirsthead
\toprule
Field & Description\\
\midrule
\endhead
\bottomrule
\endfoot
\texttt{evidence\_id} & Unique record identifier.\\
\texttt{source} & Article-level metadata, including DOI, title, journal, year, authors, citation count and local source path when available.\\
 \texttt{pico} & Structured study context, including population, intervention, comparison and outcome metrics.\\
\texttt{core\_entities} & Extracted biomedical entities with raw names, semantic classes and normalized identifiers when available.\\
\texttt{target\_disease} & Disease context for the resource: HCC, CRC or SLE.\\
\texttt{bio\_mechanism} & Text field describing the reported biological mechanism when available.\\
\texttt{phenotype} & Text field describing the reported phenotype or clinical outcome when available.\\
\texttt{experimental\_context} & Short description of the model, assay, patient group or trial context.\\
\texttt{study\_design} & Study type label, such as cohort, randomized trial, in vivo, in vitro, meta-analysis, case-control or computational.\\
\texttt{clinical\_stage} & Translational stage label, such as preclinical, clinical, phase I, phase II, phase III or phase IV.\\
\texttt{statistics} & Quantitative attributes extracted from source text, including $p$-value, sample size, fold change, confidence interval, effect size and statistical method when reported.\\
\texttt{score} & Computational evidence-quality fields, including component scores, composite score and grade level.\\
\texttt{source\_text} & Supporting text span from which the evidence record was derived.\\
\texttt{extraction\_confidence} & Model-reported confidence for the extracted record.\\
\texttt{linked\_tarkg\_entities} & Normalized entity links used to connect the record to the graph entity layer.\\
\texttt{evidence\_relations} & Directed relations to other evidence records with relation type, similarity score and rationale.\\
\texttt{merged\_from} & Identifiers of records absorbed during duplicate handling.\\
\texttt{review\_status} & Review state. In the machine-generated release this field is retained to support future curation.\\
\texttt{created\_at}, \texttt{updated\_at}, \texttt{version} & Lifecycle and version metadata.\\
\end{longtable}

At the record level, the fields fall into four main groups. The \texttt{source}, \texttt{source\_text} and \texttt{pico} objects preserve provenance and local study context. The \texttt{core\_entities} and \texttt{linked\_tarkg\_entities} fields preserve raw biomedical mentions together with normalized identifiers when available. The \texttt{statistics}, \texttt{study\_design}, \texttt{clinical\_stage}, \texttt{score} and related metadata fields describe quantitative and quality-related properties of the extracted observation. The \texttt{evidence\_relations}, \texttt{merged\_from} and version fields document how the record participates in duplicate handling, relation annotation and release tracking. A tab-separated field dictionary in the deposited \texttt{schema/} directory provides the same field names and short descriptions in machine-readable form. Missing values are represented as empty strings, empty lists or null values depending on field type. Missing values indicate that the information is not available in the extracted text or is not assigned during construction, and they should not be interpreted as negative biomedical evidence.

The companion graph files use a NetworkX node-link JSON representation with top-level release metadata, node records and edge records. Evidence nodes are derived from the deposited evidence records, and entity nodes are derived from normalized biomedical entities linked from those records. The graph edges consist of evidence-to-entity links and evidence-to-evidence relations. Figure~\ref{fig:data_overview_global} provides a compact structural view of these companion files. Because the graph layer preserves \texttt{evidence\_id} values, graph navigation can return to the corresponding source-grounded record whenever source text, quantitative detail or article metadata need to be inspected. The graph files thus serve as derived views of the primary record files.

\begin{figure}[!tbp]
\centering
\includegraphics[width=\textwidth,height=0.78\textheight,keepaspectratio]{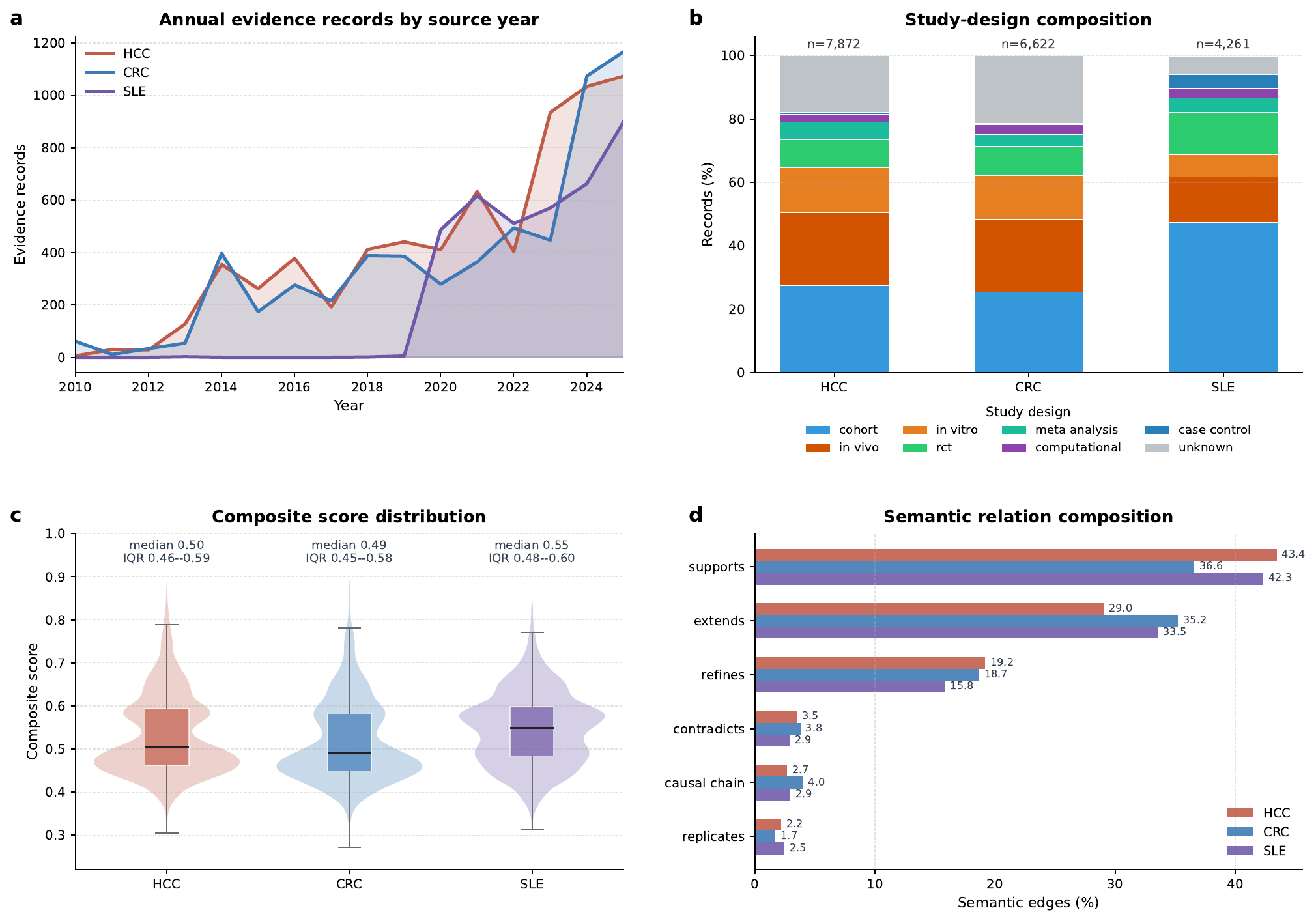}
\caption{Statistical overview of the three released disease subsets. Panel A shows the annual distribution of source years for released evidence records. Panel B shows the study-design composition of each disease subset. Panel C shows the distribution of composite evidence scores. Panel D shows the composition of released evidence-relation annotations associated with the companion graph layer.}
\label{fig:data_overview_statistics}
\end{figure}

Record counts, graph-node counts, graph-edge counts, field coverage summaries and validation outputs reported in this manuscript are generated from the same underlying release files prepared for deposition. This alignment between primary records, companion graph files and release documentation keeps the provenance of the published dataset transparent and auditable. The three deposited subsets differ in corpus size and record composition while remaining structurally aligned under the same schema. HCC contains 7,872 evidence records linked to a companion graph with 10,328 nodes and 49,756 edges, CRC contains 6,622 records with 8,795 nodes and 39,361 edges, and SLE contains 4,261 records with 6,342 nodes and 21,645 edges. Across the three subsets, source years span similar recent windows, study-design labels are heterogeneous, composite evidence scores occupy overlapping but non-identical ranges, and evidence-relation annotations remain available in the companion graph layer. Figure~\ref{fig:data_overview_statistics} provides a compact release-level summary of these deposited distributions.

\section{Technical Validation}

Technical validation evaluates whether the deposited records remain traceable to source text, sufficiently populated for reuse and internally consistent across the released files. The validation is organized around release-consistency audits, field-completeness summaries, targeted manual review of sampled components, external concordance analysis against curated disease knowledge and a record-lookup assessment.

\subsection{Schema and release consistency checks}

All three disease resources contain the same top-level record fields and graph-file structure. The graph edge totals in this manuscript use the \texttt{edges} list in each \texttt{evidence\_graph.json} file as the authoritative edge count, avoiding any mixture of filtered visualization edges and complete graph edges.

Graph consistency is checked by treating the record files as the source of truth. For each disease subset, all evidence identifiers present in the record-level JSON file are compared with the evidence-node identifiers present in the graph file, and each graph edge is checked to confirm that both endpoints appear in the graph node list. These checks detect missing evidence nodes, orphan graph evidence nodes and graph edges pointing to absent nodes.

Across all three disease subsets, these consistency checks find no missing edge endpoints, no evidence records absent from the companion graph files and no graph evidence nodes without corresponding record-level entries. The released graph files therefore match the primary record files at the level of evidence identifiers and edge endpoints.

\subsection{Field completeness checks}

\begin{table}[ht]
\centering
\caption{Coverage of selected evidence-record fields.}
\label{tab:field_coverage}
\small
\begin{tabular}{lccc}
\toprule
Field & HCC & CRC & SLE \\
\midrule
Comparison & 4492 (57.1\%) & 3951 (59.7\%) & 2598 (61.0\%) \\
p-value & 800 (10.2\%) & 835 (12.6\%) & 571 (13.4\%) \\
Sample size & 1955 (24.8\%) & 1856 (28.0\%) & 1648 (38.7\%) \\
Fold change & 340 (4.3\%) & 325 (4.9\%) & 323 (7.6\%) \\
Bio-mechanism & 5355 (68.0\%) & 4585 (69.2\%) & 2344 (55.0\%) \\
Phenotype & 7666 (97.4\%) & 6480 (97.9\%) & 4148 (97.3\%) \\
Experimental context & 7872 (100.0\%) & 6622 (100.0\%) & 4261 (100.0\%) \\
Source text & 7872 (100.0\%) & 6622 (100.0\%) & 4261 (100.0\%) \\
\bottomrule
\end{tabular}
\end{table}

Table~\ref{tab:field_coverage} reports the proportion of records in each deposited subset for which a selected field is populated. The coverage profile shows three expected patterns. First, provenance and reuse-critical fields such as \texttt{experimental\_context} and \texttt{source\_text} are complete across all three disease subsets, which supports traceability back to the original passage. Second, high coverage of \texttt{phenotype} and moderate-to-high coverage of \texttt{bio\_mechanism} indicate that most records preserve a usable description of the reported finding even when mechanistic detail is uneven across the literature. Third, lower coverage for quantitative attributes such as \texttt{p-value}, \texttt{sample size} and \texttt{fold change} reflects information that is often not explicitly reported in the source passage or not tightly attributable to the extracted observation. This conservative convention avoids backfilling values from broader article context when a numerical attribute is not directly associated with the evidence record.

\subsection{Validation design across record components}

The released fields are validated with checks matched to their role in the dataset. Provenance fields are reviewed for bibliographic plausibility and traceability, source-grounded finding fields against the supporting passage, study-context labels against explicit textual cues, entity links against auditable mention grounding, and fusion or relation fields for record coherence and endpoint validity. Table~\ref{tab:field_validation_design} summarizes these component-specific checks, and the corresponding audit results are presented in the validation subsections that follow.

\begin{longtable}{>{\raggedright\arraybackslash}m{4.2cm}>{\raggedright\arraybackslash}m{10.0cm}}
\caption{Validation checks used for the main record components.}
\label{tab:field_validation_design}\\
\toprule
Record component & Validation criterion\\
\midrule
\endfirsthead
\toprule
Record component & Validation criterion\\
\midrule
\endhead
\bottomrule
\endfoot
\texttt{source} metadata & DOI, title and publication year are checked for plausibility, and obvious metadata outliers are reviewed against article and repository information.\\
\texttt{source\_text} & Sampled records are checked for direct support from the deposited source passage.\\
\texttt{pico}, \texttt{phenotype}, \texttt{bio\_mechanism} & Sampled records are checked for reusable, source-grounded findings rather than background text alone.\\
\texttt{study\_design}, \texttt{clinical\_stage} & Study-context labels are checked against explicit wording or unambiguous cues in the source passage.\\
\texttt{core\_entities}, \texttt{linked\_tarkg\_entities} & Accepted links are checked when the mention can be grounded in the record context, focusing the audit on deposited normalized identifiers.\\
\texttt{merged\_from}, \texttt{version} & Sampled merged records are reviewed for coherence of the retained canonical record after duplicate handling.\\
\texttt{evidence\_relations} & Relation endpoints are checked for validity, and sampled links are reviewed for semantic-label consistency with record content.\\
Companion graph files & Record-to-graph and graph-to-record consistency checks confirm that the derived graph files match the primary record files.\\
\end{longtable}

As part of provenance QC, bibliographic outliers are checked when publication-year or identifier metadata appear implausible. In this release, manual review resolves a small set of SLE metadata discrepancies, and no pre-1990 metadata-year outliers remain in the released records after review. The released QC documentation preserves the affected evidence identifiers together with the corrected bibliographic fields and review status.

\subsection{Targeted manual audit}

A balanced structured audit is performed across HCC, CRC and SLE. The audit covers 150 evidence-extraction cases, 150 grounded entity-linking cases, 90 duplicate-fusion cases and 150 evidence-relation cases. Sampling is stratified by disease and component type so that the reviewed set spans multiple study designs, entity classes, relation types and merge scenarios, and the released \texttt{validation/} files preserve the structured case sheets and labels used for this audit. Entity-linking validation is restricted to high-confidence links whose raw or normalized entity names are directly checkable in the record context, allowing the deposited identifiers to be inspected directly against the released record content. The same sampled review also supports the study-design, clinical-stage and relation-type checks summarized in the paragraph after Table~\ref{tab:component_validation}.

\begin{table}[ht]
\centering
\caption{Balanced manual audit of core record-quality components.}
\label{tab:component_validation}
\small
\begin{tabular}{lccc}
\toprule
Validation item & HCC & CRC & SLE \\
\midrule
Source-text support & 47/50 (94.0\%) & 46/50 (92.0\%) & 48/50 (96.0\%) \\
Key-entity capture & 42/50 (84.0\%) & 40/50 (80.0\%) & 45/50 (90.0\%) \\
Grounded entity-link mention & 50/50 (100.0\%) & 50/50 (100.0\%) & 50/50 (100.0\%) \\
Normalized identifier validity & 50/50 (100.0\%) & 50/50 (100.0\%) & 50/50 (100.0\%) \\
Fusion record coherence & 25/30 (83.3\%) & 29/30 (96.7\%) & 28/30 (93.3\%) \\
Evidence-relation edge validity & 50/50 (100.0\%) & 50/50 (100.0\%) & 50/50 (100.0\%) \\
\bottomrule
\end{tabular}
\end{table}

Table~\ref{tab:component_validation} reports the number of sampled records or links that satisfy the core audit checks most directly tied to record usability. The highest agreement is observed for source grounding, identifier validity and edge validity, with somewhat lower agreement for key-entity capture and fusion coherence.

Additional review of the more interpretive annotation layers shows lower agreement for study-design labels (58.0\% in HCC, 74.0\% in CRC and 72.0\% in SLE) and for evidence-relation type consistency (68.0\% in all three diseases). These labels are harder to assign than source grounding or identifier validity because many biomedical articles mix preclinical, translational and clinical elements within the same paper, and relation boundaries such as SUPPORTS, EXTENDS and REFINES often depend on contextual judgment rather than explicit lexical markers. Source provenance, source-text grounding, graph integrity and normalized identifier validity remain the strongest anchors in the released files for audit and retrieval, while study-design labels, duplicate-fusion outputs and relation-type annotations function more appropriately as interpretive filtering or review aids to be read alongside the linked source passage rather than as definitive endpoint labels in isolation.

\subsection{External concordance with curated disease knowledge}

To assess whether the released resource recovers disease-relevant entities already represented in an external curated knowledge base, we compare high-support normalized entities from EvidenceNet with disease-specific entries from the Open Targets Platform API\cite{open_targets} as queried in May 2026. For each disease, we extract the 30 most frequently linked normalized gene entities and the 20 most frequently linked normalized drug entities from the deposited graph files. These lists are then compared with the top 600 disease-associated targets returned by Open Targets and with the disease-specific list of clinical candidates and approved drugs.

\begin{table}[ht]
\centering
\caption{External concordance between high-support EvidenceNet entities and curated disease knowledge from Open Targets.}
\label{tab:external_concordance}
\small
\begin{tabular}{>{\raggedright\arraybackslash}m{1.4cm}>{\raggedright\arraybackslash}m{2.7cm}>{\raggedright\arraybackslash}m{2.7cm}>{\raggedright\arraybackslash}m{6.0cm}}
\toprule
Disease & Gene overlap & Drug overlap & Representative concordant entities \\
\midrule
HCC & 17/30 (56.7\%) & 16/20 (80.0\%) & Genes: CD274, CTLA4, GPC3, EGFR. Drugs: sorafenib, lenvatinib, cabozantinib, regorafenib. \\
CRC & 16/30 (53.3\%) & 13/20 (65.0\%) & Genes: BRAF, EGFR, APC, TP53. Drugs: cetuximab, oxaliplatin, irinotecan, capecitabine. \\
SLE & 14/30 (46.7\%) & 13/20 (65.0\%) & Genes: CCL2, CD19, TLR7, CD40LG. Drugs: hydroxychloroquine, cyclophosphamide, mycophenolate mofetil, azathioprine. \\
\bottomrule
\end{tabular}
\end{table}

Table~\ref{tab:external_concordance} shows that a substantial fraction of the most frequently linked disease-specific genes and drugs in EvidenceNet are also recovered from an external curated resource. The drug concordance is particularly strong, with 65.0--80.0\% of the top 20 disease-linked drug entities overlapping Open Targets clinical candidates or approved drugs. Gene concordance is more moderate, which is expected because EvidenceNet preserves literature-grounded disease-specific evidence while the external reference emphasizes curated disease-associated targets. Overall, the concordant entities provide independent support for the biological relevance of the released normalized entity layer.

\subsection{Internal source-record lookup consistency}

To assess whether the deposited fields are sufficient for source-oriented lookup, 75 held-out prompts are generated for each disease from record intervention, comparator when available and phenotype fields. Records from the same disease subset are then ranked with deterministic token-overlap matching, and lookup quality is summarized using Top-1, Top-5, Top-10 and mean reciprocal rank. Because the prompts are derived from deposited record content, the analysis is designed as an internal source-record consistency check within the release rather than as an external retrieval benchmark.

\begin{table}[ht]
\centering
\caption{Source-record lookup from held-out record-derived prompts.}
\label{tab:qa_retrieval}
\small
\begin{tabular}{lrrrrr}
\toprule
Disease & Questions & Top-1 & Top-5 & Top-10 & MRR \\
\midrule
HCC & 75 & 0.973 & 0.987 & 1.000 & 0.982 \\
CRC & 75 & 0.973 & 1.000 & 1.000 & 0.987 \\
SLE & 75 & 0.987 & 0.987 & 0.987 & 0.988 \\
\bottomrule
\end{tabular}
\end{table}

Table~\ref{tab:qa_retrieval} reports whether each held-out source record is recovered from compact prompts derived from the same released record layer. The consistently high Top-\(k\) and MRR values indicate that the deposited fields preserve sufficient lexical and structured context for source-record lookup within the release. These values should be interpreted as evidence of internal release consistency rather than as benchmark performance on an independent biomedical retrieval or question-answering task.

\section{Usage Notes}

Record-level JSON files are the primary entry point for inspecting evidence provenance, source text and structured study context. The graph files are companion representations for network loading, visualization and linked record inspection, and the \texttt{evidence\_id} field links graph evidence nodes back to the corresponding full record. A typical workflow is to select a disease subset, filter records by fields such as \texttt{study\_design}, \texttt{clinical\_stage}, \texttt{score.grade} or normalized entity identifiers, and then inspect \texttt{source}, \texttt{source\_text}, \texttt{pico} and \texttt{statistics}.

For retrieval-oriented reuse, the complete evidence record is the recommended indexing unit. Fields such as \texttt{pico}, \texttt{phenotype}, \texttt{bio\_mechanism}, \texttt{experimental\_context}, \texttt{core\_entities} and \texttt{source\_text} can be concatenated or indexed separately, while \texttt{evidence\_id} should be retained so that retrieved items remain traceable to the deposited record and article metadata. The companion graph files can be loaded with node-link readers such as NetworkX after reading the nested \texttt{graph} object.

This release does not exhaust the HCC, CRC or SLE literature. The records are machine-generated and validated on balanced samples rather than on every field in every record. Score and grade fields are most useful for filtering and prioritization, whereas normalized entity links, study-design labels and evidence-relation labels are best interpreted alongside the linked source passage. Lower coverage of quantitative fields indicates that a value was not directly extractable or not confidently attributable to the record. Original full-text PDFs are not redistributed with the release; instead, the deposited \texttt{source\_text} fields retain limited evidence-bearing passages for provenance inspection, and users should consult the cited source articles through DOI, PubMed or publisher access for full-document context. Example loading and validation scripts are provided in the accompanying code repository, and record and graph files should be used from the same deposited release.

\section{Data Availability}

The EvidenceNet HCC, CRC and SLE evidence-record dataset is available from Figshare at \url{https://doi.org/10.6084/m9.figshare.31888399}. The release includes \texttt{records/}, \texttt{graphs/}, \texttt{schema/}, \texttt{validation/}, \texttt{prompts/} and \texttt{checksums\_sha256.txt}. The record-level JSON files are the primary data records and the graph JSON files are companion representations. The release does not redistribute the original article PDFs; it instead preserves article metadata, DOI-level provenance and limited supporting source-text passages for audit and reuse.

\section{Code Availability}

The construction and validation code is available from the EvidenceNet GitHub repository at \url{https://github.com/ZUST-BIT/EvidenceNet-code} and from the versioned Zenodo archive at \url{https://doi.org/10.5281/zenodo.19748584}. The code package contains scripts for PDF parsing, evidence extraction, entity normalization, evidence scoring, duplicate handling, graph construction, visualization, metadata correction and validation, together with the configuration files used for the released workflow. The archived configuration for the deposited release records the primary OpenAI extraction model (\nolinkurl{GPT-5.1}), the lighter auxiliary model (\nolinkurl{GPT-5-mini}), temperature-zero inference and the main similarity thresholds used during duplicate handling and relation construction.

\section{Author Contributions}

Chang Zong: conceptualization, methodology, software, data curation, formal analysis, validation, visualization, writing--original draft. Sicheng Lv: data curation, software, formal analysis, writing--review and editing. Si-tu Xue: investigation, validation, writing--review and editing. Huilin Zheng: data curation, validation, visualization, writing--review and editing. Jian Wan: supervision, methodology, writing--review and editing. Lei Zhang: conceptualization, supervision, project administration, funding acquisition, writing--review and editing. All authors approved the final manuscript.

\section{Competing Interests}

The authors declare no competing interests.

\section{Funding}

This work was supported by the ``Pioneer'' and ``Leading Goose'' R\&D Program of Zhejiang (Key Research and Development Program of Zhejiang Province), China (Grant No. 2025C01115).

\section{Ethics Statement}

No new human participant data, animal experiments or clinical specimens are collected for this dataset. The deposited records are derived from published articles and retain article-level provenance and source-text passages for review. The dataset does not include individual-level patient records beyond information already present in the source publications.

\end{document}